\documentclass{article}
\usepackage[letterpaper,top=2cm,bottom=2cm,left=3cm,right=3cm,marginparwidth=1.75cm]{geometry}
\usepackage[colorlinks=true, allcolors=blue]{hyperref}
\usepackage{authblk}

\usepackage{siunitx}
\usepackage{booktabs}

\usepackage[english]{babel}
\usepackage{amsmath}
\usepackage{graphicx}
\usepackage{xcolor}
\usepackage[toc,page]{appendix}

\usepackage[
   backend=bibtex,
   style=numeric-comp,
   sorting=none,
]{biblatex}
\usepackage{csquotes}
\AtEveryBibitem{%
\ifentrytype{article}{
    \clearfield{url}%
    \clearfield{urlyear}%
    \clearfield{issn}%
    \clearfield{language}%
    \clearfield{langid}%
}{}
\ifentrytype{collection}{
    \clearfield{url}%
    \clearfield{urlyear}%
}{}
\ifentrytype{incollection}{
    \clearfield{url}%
    \clearfield{urlyear}%
}{}
}

\begin{document}

\title{Stable, bidirectional electro-optic transduction in thin film lithium tantalate}

\author[1]{Christopher J. Axline\thanks{Correspondence to: christopher.axline@miraex.com}}
\author[1]{Stephan Gamper}
\author[1]{Phoebe M. Tengdin}
\author[1]{Moritz Businger}
\author[2]{Guilhem Alma}
\author[2]{Marina A. Roquet}
\author[1]{Nicola Brusadin}
\author[1]{Robin Giroud}
\author[2]{Luis G. Villanueva}
\affil[1]{Miraex SA, EPFL Innovation Park, Building L South, Chem. de la Dent d'Oche 1B, 1024 Ecublens, Switzerland}
\affil[2]{École Polytechnique Fédérale de Lausanne (EPFL), 1015 Lausanne, Switzerland}

\maketitle
\begin{abstract}

Efficient and stable microwave–optical transduction is a key enabling technology for distributed superconducting quantum computing and heterogeneous quantum networks.
Electro-optic transducers based on thin-film lithium niobate (TFLN) have shown strong promise, but demonstrations to date have been limited by various factors such as low frequency bias drift, low efficiency, fabrication complexity, and scalability.
Here we demonstrate the first integrated electro-optic microwave–optical transducers realized in thin-film lithium tantalate (TFLT), a material platform offering Pockels nonlinearity comparable to TFLN together with improved bias stability and high-power handling.
We fabricate superconducting microwave resonators coupled to tunable photonic-molecule optical resonators using wafer-scale deep ultraviolet lithography, offering high-throughput production of hundreds of devices per wafer.
Across six devices we observe coherent bidirectional conversion between C-band optical photons and 4.9–5.5\,GHz microwave photons, with measured on-chip efficiencies and inferred single-photon coupling rates $g_0 / (2\pi) \sim 1$\,kHz consistent with theory.
Continuous operation over multiple days is achieved using a static bias field with minimal feedback, demonstrating a major operational advantage.
We further characterize optical loss statistics, microwave resonator performance, and optically induced added noise under pulsed pumping, finding less than one added photon for \SI{100}{\micro\second} pulses at the highest measured efficiencies.
These results establish TFLT as a scalable and robust electro-optic platform for future quantum interconnects and modular quantum processors.
\end{abstract}

\section{Introduction}

Quantum computing platforms are rapidly advancing, with superconducting qubits, trapped ions, and spins in solids each demonstrating key milestones toward fault-tolerant architectures \cite{acharya_quantum_2025,loschnauer_scalable_2025,team_digitally_2026}.
Modular approaches based on distributed quantum computing have been proposed as a pathway beyond monolithic laboratory-scale processors, while optical networking may further enable heterogeneous systems that interconnect distinct quantum hardware platforms \cite{kimble_quantum_2008,monroe_large-scale_2014,han_microwave-optical_2021}.

Low-loss optical fiber links can preserve quantum information over long distances and are therefore an attractive interconnect medium for modular quantum processors \cite{krutyanskiy_light-matter_2019,liu_long-lived_2026}, provided coherent interfaces between local qubits and optical photons are available.
Promising microwave frequency qubit platforms, however, would require coherent, efficient, and rapid transduction between microwave and optical photons to take advantage of optical networks.

Transduction has been demonstrated on numerous integrated platforms using a variety of approaches, including direct (single stage) and indirect (multi-stage) transduction \cite{fan_superconducting_2018,fu_cavity_2021,xu_bidirectional_2021,jiang_optically_2023,weaver_integrated_2024,zhao_quantum-enabled_2024,warner_coherent_2025,mohl_bidirectional_2025,zhou_kilometer_2025,multani_integrated_2026}.
While indirect approaches (such as piezo-opto-mechanical) often use strong couplings between each stage, intermediate modes can limit bandwidth and introduce high added noise that can be difficult to control \cite{han_microwave-optical_2021}.
Direct approaches, like electro-optic (EO) coupling using the Pockels effect, have shown comparable efficiency despite lower intrinsic coupling rates.
EO systems like thin-film lithium niobate (TFLN) have emerged as a dominant material system for this purpose, enabled in particular by high-quality optical resonators \cite{zhu_twenty-nine_2024}.
Electro-optic materials like aluminum nitride \cite{zhou_kilometer_2025} and barium titanate \cite{mohl_bidirectional_2025} have also shown promising results, but can be limited by piezo-electric effects or fabrication challenges that inhibit straightforward scaling.

Lithium tantalate ($\text{LiTaO}_3$), which shares very similar EO coefficients with lithium niobate, has emerged in recent years a promising EO material for applications including quantum transduction.
Studies have highlighted distinct material advantages including a wider bandgap, reduced susceptibility to photorefractive effects and optical damage, and superior low frequency stability \cite{wang_lithium_2024,wang_thin-film_2025,sayem_high-power_2026}.
These advantages could increase the efficiency of an EO transducer using $\text{LiTaO}_3$. 
In particular, lithium tantalate's inherently stable direct current (DC) response overcomes a longstanding limitation of $\text{LiNbO}_3$ \cite{powell_dc-stable_2024}, enabling transduction over long durations using a single, static bias field.
This avoids the need to actively compensate bias drift, simplifying device operation for scaled-up quantum computing or networking applications.

Here, we report the first realization of electro-optic microwave-optical transduction in thin-film lithium tantalate (TFLT).
We integrate high-$Q$ optical resonators with a superconducting microwave resonator and demonstrate coherent coupling between microwave and C-band optical photons by performing transduction on six devices with varied designs.
We perform both microwave-to-optical and optical-to-microwave transduction and compare efficiency with theoretical predictions, measuring results that agree well with estimated single photon coupling rates $g_0 / (2 \pi) \sim$ 1\,kHz.
We evaluate the optical platform, measuring equivalent propagation losses in TFLT as low as \SI{0.25}{dB/\cm}. 
We demonstrate device stability by continuously performing days-long transduction.   
We investigate added noise from on- and off-resonant optical pump photons, finding that coupler-scattering-dominated transduction adds less than 1~photon of noise for \SI{100}{\micro\second} pulses at the highest on-chip efficiencies approximately 0.04\%.

In this work, optical and microwave circuits are patterned exclusively using wafer-scale exposure techniques, avoiding the throughput limitations associated with electron-beam lithography (which has been used to produce most integrated EO converters to date).
Thus we fabricate hundreds of converters on a single wafer with low device-to-device variance, supporting a critical requirement for the large-scale optical interconnects needed to link future quantum processors.
Combined with stable and tunable biasing on our devices, this work establishes TFLT as a powerful alternative to EO materials like TFLN and opens new opportunities for the deployment of large-scale quantum interconnects.

\section{Device design and operation}

\begin{figure}[htbp]
    \centering
    \includegraphics[width=1\linewidth]{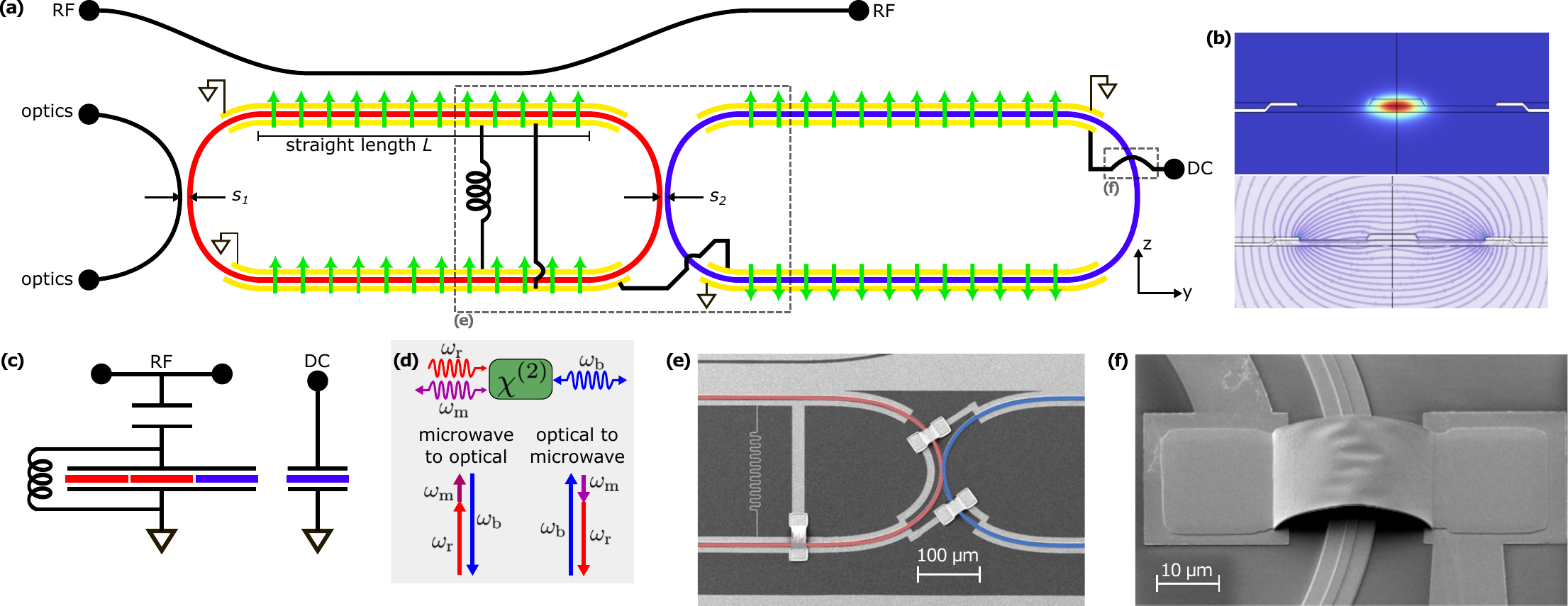}
    \caption{
    (a) In each measured transducer, coupled optical racetrack resonators (red: ``bright'' resonator; blue: ``dark'' resonator) interface with a microwave resonator (including capacitor electrodes, colored gold).
    Signals are introduced via radiofrequency (RF) and optical bus lines, whereas DC bias is routed separately to electrodes that tune the dark resonator.
    Green arrows in the EO interaction regions indicate the direction of electric field lines when positive RF or DC voltage is applied.
    Device straight lengths ($L$) is varied, as are gap distances between the bright resonator and the optical bus ($s_1$) or the dark resonator ($s_2$).
    Electrodes are connected with superconducting traces and bridges (black lines; see panel f).
    (b) Transduction is possible by overlapping electric field components within the EO material.
    Here within a cross section of the electrode--waveguide stack in the interaction region, simulated optical field magnitude (top) and electric field lines (bottom) are shown.
    (c) Electric circuit diagram of the RF resonator modeled as a parallel LC resonator capacitively coupled to the RF bus.
    The capacitor can be modeled as a parallel plate capacitor loaded by both straight sections of the bright resonator and one straight section of the dark resonator.
    A DC source applies voltage separately to the other straight section of the dark resonator.
    (d) Method of triply resonant direct electro-optic transduction between microwave and optical frequency signals.
    The EO material comprising the waveguides, TFLT, acts as a $\chi^{(2)}$ source for parametric three-wave mixing.
    An optical pump at frequency $\omega_\text{r}$ converts between optical photons at $\omega_\text{b}$ and microwave photons at $\omega_\text{m}$.
    Scanning electron micrographs show (e) false-colored bright and dark optical resonators around the resonator coupling region indicated in (a), and (f) a superconducting air bridge used to route electrical signals over the optical layers without need for an additional cladding material.
    }
    \label{fig:scheme}
\end{figure}

We implement quantum transduction using a triply resonant electro-optic architecture (Figure~\ref{fig:scheme}) based on two coupled optical racetrack resonators and a microwave resonator.
Two equal-length, evanescently coupled optical resonators composed of lithium tantalate rib waveguides form a photonic molecule \cite{zhang_electronically_2019} similar to demonstrations using lithium niobate \cite{holzgrafe_cavity_2020,xu_light-induced_2024,warner_coherent_2025}.  
The photonic molecule supports symmetric (S) and antisymmetric (AS) optical supermodes with frequencies $\omega_\text{S,AS}$ \cite{soltani_efficient_2017}.
The supermodes span both cavities, wherein the light is in (S) or out of (AS) phase.
Transduction is performed by engineering the difference in supermode frequencies such that $\omega_\text{S} - \omega_\text{AS} \approx \omega_\text{m}$
where $\omega_\text{m}$ is the microwave resonator frequency, realized by a quasi-lumped-element LC superconducting resonator.
The triply resonant scheme avoids the need for $\omega_\text{m}$ to match the optical resonators' free spectral range (FSR) while suppressing the unwanted down-converted sideband, because it is non-resonant.

The single-photon interaction rate $g_0$ and cooperativity $C_\text{eo}$ between transducer modes are derived in Appendix~\ref{app:hamiltonian} and given in Table~\ref{tab:parameters} for measured devices.
In particular, $g_0$ can be related to electrode coverage fraction $\alpha$ and the magnitude of microwave field in the $z$ direction, $E_{\text{m},z}$.
Resonator straight length $L$ varies from \SI{150}{\micro\meter} to \SI{500}{\micro\meter} in our designs, varying $\alpha$ and $|E_{\text{m},z}|$ minimally (13\% and 20\%, respectively).
Including all simulated contributions, we expect $g_0 / (2 \pi)$ to vary from 1.03--1.30\,kHz between long and short designs.
Parameter estimates are given for each measured devices in Table~\ref{tab:parameters}.

\paragraph{Resonator tuning.}

\begin{figure}[htbp]
    \centering
    \includegraphics[width=1\linewidth]{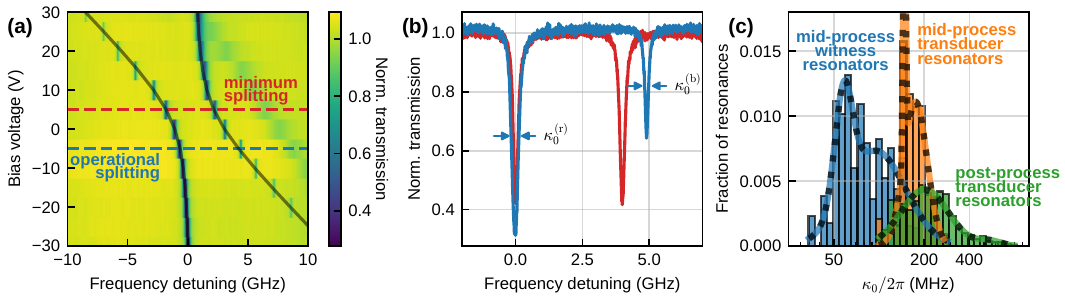}
    \caption{
    (a) Optical spectroscopy shows the locations of resonances in Device~1 near 1560\,nm as voltage is applied to bias the dark resonator.
    Dashed lines indicate bias voltages where the data in (b) are taken.
    (b) Optical transmission at the fully hybridized minimum splitting point (red line) and transduction operating point (blue line), shifted to align.
    Internal loss rates $\kappa_0$ near ``operational splitting'' bias values used in transduction are those supplied in Table~\ref{tab:parameters}.
    (c) Histograms of optical resonators' internal loss rates fit to dual log-normal distributions (dashed lines) to extract median and log-sigma values (see text) for $N=129$, $N=51$, $N=941$ resonances on mid-process witness resonators (blue), mid-process transducer resonators (orange), and post-process transducer resonators (green) respectively.
    }
    \label{fig:bias}
\end{figure}

The ``bright'' optical resonator is evanescently coupled to an optical bus, which brings light to and from grating couplers near the edge of the chip (Figure~\ref{fig:scheme}(a)).
Fiber array units (FAUs) are used to probe optical transmission at room temperature, and are then fixed using epoxy to operate cryogenically with as low as \SI{9}{\decibel} loss per facet below \SI{4}{\kelvin} (Appendix~\ref{app:packaging}).
Non-uniform material thickness and variable dimensions of fabricated waveguides can cause the two resonators to have different phases.
Dedicated DC electrodes are placed around a portion of the ``dark'' resonator to support electro-optic tuning.
Applying a voltage can tune the resonator's phase to match that of bright resonator, forming a fully hybridized photonic molecule displaying an avoided crossing spectroscopically (Figure~\ref{fig:bias}(a)).
The dark resonance can be detuned blue or red of the bright resonance, but for transduction to take place, the pump must be applied on the red-detuned mode such that $\omega_\text{S} = \omega_\text{r}$.
The optical signal thus passes through the mode $\omega_\text{AS} = \omega_\text{b}$.

In our devices, device loss and coupling parameters that would lead to optimum transduction typically occur at the ``minimum splitting'' condition, where the distribution of energy in the supermodes is equal between the resonances.
To efficiently transduce, however, this splitting must match the microwave resonator frequency (``transduction condition'') within the transduction bandwidth, typically $<20$\,MHz.
The coupled resonator transmission spectrum is fit in both cases (Figure~\ref{fig:bias}(b)) and the results presented in Table~\ref{tab:parameters}.
    
\paragraph{Design of optical and microwave resonators.}

Our chosen coupled racetrack resonator design was previously demonstrated in TFLN \cite{holzgrafe_cavity_2020}, serving as a benchmark against which we can compare the same design on our new TFLT platform.
Waveguide width and the size of the gap between electrodes are designed to balance propagation loss (simulated to be limited by electrodes to \SI{<0.28}{dB/\meter}) with microwave field density.
Evanescent coupling between the bright and dark resonators depends the size of gap $s_2$, leading to minimum splitting values 3.4--6.2\,GHz.
Further details of transducer design are given in Appendix~\ref{app:hamiltonian}.

Microwave signals are capacitively coupled from a coplanar waveguide (CPW) transmission line as depicted in Figure~\ref{fig:scheme}(c).
We typically obtain internal quality factors $300 < Q_\text{int} < 1100$ and external (coupling) quality factors $10^{3} < Q_\text{ext} < 10^{4}$ at moderately low microwave signal powers.

\paragraph{Optical resonator performance.}

Since many device parameter requirements must be simultaneously satisfied to enable efficient transduction, it is important that device parameters can be consistently achieved.
Measurements of optical resonator parameters, and in particular intrinsic loss (which plays a significant role in transduction performance), can be measured without packaging or cooling and used to collect statistics.
We measure transducer optical resonators at a mid-process stage (prior to deposition of the superconducting layer) and after complete processing, as well as more compact circular ring resonators that serve as process witnesses (Appendix~\ref{app:resonators}).
Wafer-scale measurements show tightly clustered resonator losses, with post-process transducer modes centered near \SI{229}{\mega\hertz} (Figure~\ref{fig:bias}(c)).
Optical resonator parameters for the measured transducers are given in Table~\ref{tab:parameters}.

\paragraph{Fabrication.}

Transducers are fabricated at wafer scale using stepper based DUVL on a commercially available stack of \SI{400}{\nano\meter} of X-cut lithium tantalate (LT) on top of \SI{2}{\micro\meter} of SiO$_{2}$ and \SI{525}{\micro\meter} of Si (see Appendix~\ref{app:fabrication} for further details.).
Optical waveguides are patterned to produce \SI{200}{\nano\meter} thick LT rib waveguides on top of a patterned LT slab layer.
A NbN film is patterned using DUVL and a lift-off process to avoid damage to optical structures by etching.
The superconducting film overlaps and directly contacts the LT slab over about \SI{1}{\micro\meter} near the interaction regions as shown in Figure~\ref{fig:scheme}(b), which should concentrate the microwave electric field to improve $g_0$ by a small factor \cite{warner_coherent_2025}.

After optical and superconducting layers are patterned, superconducting bridges are produced and the chips are singulated.
Wafer-scale optical measurements are performed after major processing steps (partial results shown in Figure~\ref{fig:bias}(c)) to determine the impact of each fabrication step on the optical losses and identify any damage.

The optical waveguides are fully ``air cladded'' to avoid optical cladding materials, typically PECVD-grown oxides with high loss tangents \cite{oconnell_microwave_2008} that reduce microwave resonator performance.
With no cladding, we use superconducting air bridges \numrange[range-phrase=--]{25}{45}\,\unit{\micro\meter} long to stitch parts of the circuit together (Figure~\ref{fig:scheme}(f)).
Independent resonator measurements suggest that these bridges do not limit internal quality factors (Appendix~\ref{app:microwave}).

\section{Microwave--optical transduction}

\subsection{Transduction efficiency}

We apply a strong optical pump near 1560\,nm to characterize transduction in both directions using a setup further described in Appendix~\ref{app:setup}.
To observe optical--microwave transduction we modulate the pump with an optical intensity modulator at a frequency near $\omega_\text{m}$ and measure the resulting microwave signal on a vector network or spectrum analyzer (for CW or pulsed operation, respectively).
In microwave--optical transduction, a microwave signal near $\omega_\text{m}$ is applied to the transducer and the optical output, which includes both pump leakage and the optically transduced signal, is directed onto a photodetector, and the beatnote signal extracted. 
Due to setup limitations, bidirectional transduction is only measured on Device~1.

\begin{figure}[htbp]
    \centering
    \includegraphics[width=1\linewidth]{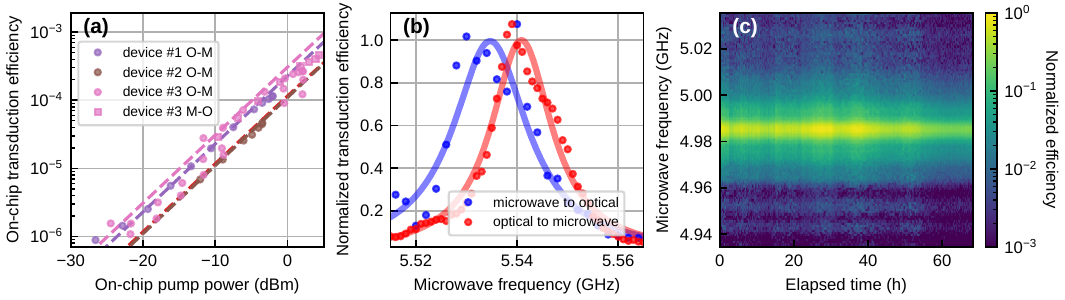}
    \caption{
    (a) Dependence of on-chip transduction efficiency on pump power for the most efficient devices 1, 2, and 3.
    Optical--microwave efficiency (``O-M'', circles) and microwave--optical efficiency (``M-O'', squares) are overlaid with predicted values from theory (dashed lines), calculated using independent measurements of microwave and optical device parameters.
    We observe the expected linear dependence.
    (b) Normalized efficiency in each transduction direction for Device~1 is shown relative to microwave frequency, with a Lorentzian fit giving a \SI{3}{\decibel} bandwidth of $15 \pm 1$\,MHz in both directions.
    (c) Normalized efficiency measured as a function of microwave frequency in Device~6 over several consecutive days.
    The transducer is operated in CW mode.
    Slow feedback is applied to track laser frequency, and intensity changes due to polarization are tracked and accounted for.
}
    \label{fig:efficiency}
\end{figure}

On-chip transduction efficiency is observed to closely follow theoretical predictions (Figure~\ref{fig:efficiency}(a)), which are derived from independent measurements of microwave and optical frequencies and internal and external losses scaled by an inferred value of $g_0$.
This results in $g_0/(2\pi)$ as high as \SI{1.1}{\kilo\hertz} (in Devices~1 and 4), which agrees well with predicted value $\sim 1$\,kHz.

Transduction efficiency was measured as a function of the microwave frequency $\omega_\text{m}$, in both transduction directions, finding a process bandwidth near \SI{15}{\mega\hertz} in both cases (Figure~\ref{fig:efficiency}(b)).
While peak efficiency was similar in both directions, the values of $\omega_\text{m}$ at peak efficiency differed by about \SI{6}{\mega\hertz}, which is significant relative to the bandwidth.
This could indicate detuning of experimental parameters, like bias voltage, away from the optimum, in particular for the microwave--optical transduction direction where maximum detected signal may be obtained away from resonance under certain conditions.

We further observed that the sensitivity of splitting shift of the coupled optical resonators to applied bias voltage, which should linearly relate to field magnitude between the capacitor plates, reduces by about 50\% when cooling from room temperature to $<4$\,K (Table~\ref{tab:parameters}).
This reduction is similar in magnitude to that reported in LN and other electro-optic materials \cite{herzog_electro-optic_2008,thiele_cryogenic_2022,bisang_plasmonic_2024,ulrich_engineering_2025,mohl_bidirectional_2025}.
If this change originates from temperature dependence of the material's electro-optic $r_{33}$ parameter (for example through mechanically mediated stresses), then $g_0$ would nearly double if this effect could be avoided.
The shortest resonators, which should have $g_0$ about 30\% larger than the longest devices, were not measured, but among the lengths measured, there was no significant correlation between $g_0$ and length. 

While operating transduction at cryogenic temperatures as low as \SI{0.1}{\kelvin}, we monitored the dependence of transduction efficiency on bias voltage to check for bias drift.
We generally observed that pulsed transduction could be operated stably for hours, or even days, with minimal feedback (Figure~\ref{fig:efficiency}(c)).
We attributed small differences observed over time to laboratory conditions outside of the cryogenic device setup.
Prolonged application of high on-chip optical powers (such as constant powers above \SI{-20}{dBm}), however, seemed to accelerate bias drift, possibly through charge generation and redistribution through photorefractive, photocharging, and thermal effects \cite{shen_parasitic_2024}.

\subsection{Added noise}

When applying high optical pump energies, heightened quasiparticle density is known to degrade the microwave resonator's performance \cite{mohl_bidirectional_2025}.
We typically observe a saturation of transduction efficiency above about \SI{-20}{dBm} of continuous on-chip optical power, corresponding with a decrease in the resonance frequency and internal quality factor of the microwave resonator (Figure~\ref{fig:shifts}).
This effect can be avoided by operating at lower powers or by driving with short optical pulses.

We typically operate transduction, as well as noise characterization, in a pulsed fashion by gating the optical pump and sideband with an acousto-optic modulator (AOM).
In this configuration, higher optical powers can be introduced to the chip while keeping average power low, commensurate with the duty cycle of the pulsing.

In optical--microwave transduction, thermal input noise from the optical signal should be negligible, and so optically induced microwave noise should dominate the output noise signal \cite{han_microwave-optical_2021}.
We observe optically induced noise by omitting the optical sidebands in a setup otherwise configured to perform \emph{optical--microwave} transduction.
The output signal is amplified through a measurement chain whose gain is independently calibrated and then measured on a spectrum analyzer to determine a noise rise.

\begin{figure}[htbp]
    \centering
    \includegraphics[width=1\linewidth]{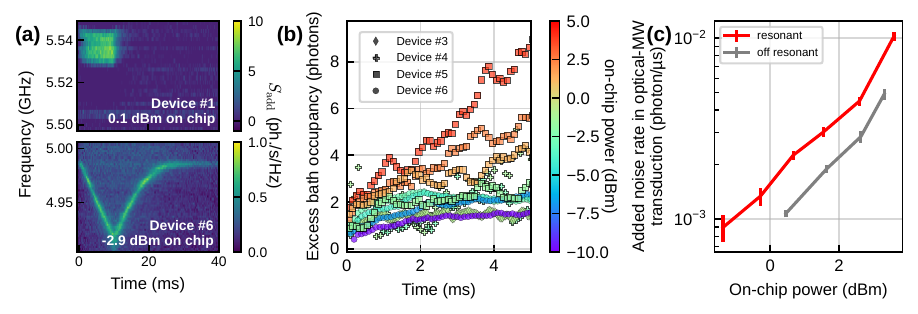}
    \caption{
    (a) Added noise signal for Device~1 (top) and Device~6 (bottom), given as a rate of measured excess photons per bandwidth, for a \SI{10}{\milli\second} optical pulse differ in the magnitude of detected noise as well as each resonator's frequency shift.
    Though Device~6 has a \SI{7}{\mega\hertz} bandwidth that is smaller than the \SI{27}{\mega\hertz} bandwidth of Device~1, it shifts by nearly \SI{80}{\mega\hertz} in \SI{10}{\milli\second}, while Device~1 shifts by less than \SI{5}{\mega\hertz}. 
    (b) The photon number occupancy above the device temperature is inferred from detected added noise and device parameters, for different devices (symbols) and optical powers (colors).
    Occupancy at $t=0$ appears to depend on power for this configuration, with a 2.5\% duty cycle of applied optical power, suggesting that the resonator is not completely thermalized to the physical device temperature between pulses.
    (c) A rate of added noise is obtained on Device~1 by performing a linear fit to noise signal in time prior to the saturation regime, for different powers of an optical pump aligned to the red resonance feature (red trace) or far detuned from it (gray trace).
}
    \label{fig:noise}
\end{figure}

When applying strong optical pumps, the output noise signal signifies a microwave resonator bath temperature exceeding the physical device temperature.
Using the thermal bath model described in Appendix~\ref{app:noise}, and knowledge of initial resonator parameters, we obtain resonant frequency, quality factors, and bath populations.
Added noise increases and resonator frequency drops as optical pulses are applied for longer periods of time (Figure~\ref{fig:noise}(a)).
For most devices, frequency shifts less than their total bandwidth (for pulses up to \SI{10}{\milli\second} and at the highest on-chip optical powers, about \SI{5}{dBm}), but Device~6 showed a stronger frequency shift.
Measured bath noise increases linearly with time and then appears to saturate in some cases (Figure~\ref{fig:noise}(b)).
Added noise depends very strongly on the level of on-chip optical power: around \SI{5}{dBm}, added noise continues to climb beyond \SI{5}{\milli\second}; around \SI{-5}{dBm}, added noise saturates in less than \SI{1}{\milli\second}; around \SI{-15}{dBm}, no added noise can be detected.

Detuning the optical pump away from the red optical resonance in Device~1 --- where we expect 7\% (0\%) of fiber-incident light to be scattered on (off) resonance --- reduces the rate at which added noise is generated by about 35\% (Figure~\ref{fig:noise}(c)).
This suggests that scattered light from grating couplers dominates added noise generation among other contributions, like on-chip waveguides and resonators.

\section{Discussion}

This first demonstration of electro-optic transduction in an integrated TFLT devices establishes TFLT as a strong alternative to TFLN to construct quantum transducers that are more stable, versatile, and scalable.
The stable bias fields we observed enable long term operation without feedback.
While unstable bias has previously been corrected with swept bias voltages \cite{holzgrafe_cavity_2020}, poly-Si layers \cite{shen_parasitic_2024}, or material treatments such as annealing \cite{holzgrafe_relaxation_2024}, these approaches inhibit scaling compared to using intrinsically stable electro-optic material.
Our wafer-scale, DUVL-based transducer microfabrication approach, as well as more compact designs enabled by LT's low anisotropy, allows higher device yield per run than is possible with e-beam lithography or TFLN based approaches.
This scalability paves the way for integrated multi-channel converters that may be required to distribute a sufficient quantity, fidelity, and/or rate of entanglement resources within a quantum network.

\paragraph{Cryogenic performance of LT.}

When cooled, our transducers show a consistent reduction in apparent $r_{33}$ at cryogenic temperatures as previously reported.
It could arise from temperature-dependent clamping and differential thermal stress between the TFLT, buried oxide, substrate, and metal stack that modifies the effective electro-optic tensor, and be addressed by future studies and design choices involving different material cuts or processing or design choices aimed at reducing differential material stress.
In two exceptions to this (Device~4 and 2), cryogenic $r_{33}$ is further reduced; we attribute this to collapsed wire bonds or superconducting bridges, partial short circuits from metal debris, or other electrical circuit anomalies that reduce the electric field applied to the TFLT resonator.
Moving to separable bias lines, which here are each shared among six devices, would make such a failure less likely.
Despite the reduced electro-optic response of TFLT at cryogenic temperatures, extracted values of $g_0$ remain high.

\paragraph{Microwave performance.}

Relative to previous demonstrations of TFLN racetrack resonators, we have realized several improvements in process and material stack design to support a higher $g_0$.
Most notably, we fully omit cladding oxide and fully etch away LT in all regions away from waveguides, not just near the interaction regions.
This avoids oxides grown using PECVD techniques, with loss tangents typically $> 2.7 \times 10^{-3}$ at millikelvin temperatures and single-photon excitation levels \cite{oconnell_microwave_2008}, and should reduce the likelihood to piezoelectrically excite parasitic modes with excess LT.
Even then, we observe internal resonator quality factors that do not exceed $Q_\text{int} \sim$ 1,100.
From an energy participation analysis (Appendix~\ref{app:participation}), we expect our resonator $Q_\text{int}$ to exceed 3,500 based on our material measurements, and to exceed 100,000 based on literature reports.
Therefore it will be important to understand how our microwave quality factor is truly limited, whether by a difference between bulk and processed LT, or by another means we have not characterized.

\paragraph{Improving transducer performance.}
We expect that significant improvements to both design and material properties are possible on this platform.
More compact designs have been shown to improve $g_0$ by increasing the fraction of interaction regions and reducing resonator capacitance \cite{warner_coherent_2025}.
Our values of optical loss, which play an outsized role in transducer efficiency, are still higher than values reported for other TFLT resonators \cite{powell_dc-stable_2024,wang_lithium_2024}.
We have identified process improvements to etching and other steps that reduce sidewall roughness and resonator loss.
By adjusting the design and dimensions of optical and microwave coupling sections, we expect to increase coupling strength, further increasing transduction efficiency.
Cumulatively, we expect that these improvements will reduce optical resonator losses to $\kappa_0/(2 \pi) \sim$ \SI{35}{\mega\hertz} and improve our on-chip efficiency per µW of pump power from about \SI{1e-6}{\micro\watt^{-1}} to \SI{2}{\%/\micro\watt}.
Incorporating changes to microwave resonator design to bring $Q_\text{int}$ consistently above 1,000 will also be important, resulting in a final expected on-chip efficiency around \SI{30}{\%/\micro\watt} (applicable in the low power, linear regime).

\paragraph{Improving couplers.}

We have reported on-chip transduction efficiency throughout this work, with total (off-chip) transduction efficiency adjusted by the approximate 10\,dB optical grating coupler input or output loss.
This distinction will become significant once we reach on-chip efficiencies around 1\%.
We choose to permanently affix fiber interfaces to our chips, which eases operation and scalability but prevents optimization of coupling at cryogenic temperatures.
Future efforts will focus on optimizing these attachment techniques and improving coupler loss below 0.5\,dB per coupler.
This will likely involve alternative designs such as edge couplers, photonic wirebonds, and evanescent couplers \cite{meenehan_silicon_2014,lin_cryogenic_2023,zeng_cryogenic_2023,sulway_high-performance_2023} which have been shown to be compatible with materials like TFLT and cryogenic operation.
Our results suggest that scattered light from grating couplers has so far dominated added noise generation, but if grating coupler losses were reduced below about \SI{3}{dB}, scattering from the resonators would begin to dominate.
Addressing limitations from these new scattering sources will likely require additional on-chip structures to reflect or absorb scattered light, beyond modifications to couplers and packaging.

With these anticipated improvements, optical powers \SI{<-20}{dBm} would be present on chip.
This level of power produced negligible output noise and resonator effects in our measurements, but we can extrapolate from the added photon rate in Figure~\ref{fig:noise}(c).
Adjusting for a more balanced resonator coupling factor than the undercoupled situation measured, we would still expect \SI{<1e-7}{\micro\second^{-1}} of added noise photons at these powers.

\paragraph{Entanglement protocol.}

A primary goal of microwave--optical transducers is to generate entanglement between microwave frequency quantum devices across an optical link.
To this end we estimate the performance of transducers operating within an emitter–emitter protocol using a time-bin encoding to generate Bell states between superconducting qubits \cite{shapourian_quantum_2025}.
With the improvements to the optical platform discussed above, as well as improvements to optical and microwave resonator designs, we expect to achieve a coupling rate $g_0 / (2 \pi) \sim 6$\,kHz and optical loss rate $\kappa_0^{(\text{opt})} / (2 \pi) \sim 40$\,MHz.
Transducers operating with these parameters near \SI{50}{\milli\kelvin} are simulated to produce Bell states with 99.6\% fidelity at a rate \SI{25}{\kilo\hertz} with about \SI{300}{\micro\watt} of dissipated power.
As coupler loss is reduced further by using scalable coupling technologies with lower loss than grating or edge couplers, we anticipate the ability to generate 99.9\% fidelity states at a rate \SI{300}{\kilo\hertz} with \SI{70}{\micro\watt} of dissipation.
These estimates represent one case, but higher operating temperatures are possible, and fidelity, rate, and power dissipation can be traded off. 
This suggests that core platform improvements, based on fundamentally good TLFT platform performance, will enable useful entanglement generation rates with minimal need to include quantum memories, distillation steps, or other protocol ``overhead'' that can inhibit scalability.

\section{Conclusion}
We have demonstrated the first integrated electro-optic microwave–optical transducers based on thin-film lithium tantalate, establishing TFLT as a compelling new material platform for coherent quantum frequency conversion. By combining high-$Q$ coupled optical resonators with superconducting microwave circuits, we realize bidirectional transduction between microwave and optical C-band photons with performance consistent with theoretical expectations of single photon coupling rates $g_0 / (2\pi) \sim 1$\,kHz.

A central result of this work is the intrinsic DC bias stability of lithium tantalate devices. Unlike prior thin-film lithium niobate implementations that often require active bias compensation, our transducers operate for hours to days using a single static bias with only slow correction of external laser conditions. This operational simplicity is likely to become increasingly important as converters are scaled to multi-channel architectures.

We further show that wafer-scale deep-ultraviolet lithography can be used to fabricate large numbers of transducers with reproducible optical performance, overcoming throughput limitations associated with electron-beam lithography. Statistical measurements indicate that optical losses are already sufficiently controlled to support present transduction experiments, while clear pathways exist for substantial improvement through reduced sidewall roughness, optimized couplers, and stronger resonator coupling. In parallel, microwave quality factors remain a key limiter and motivate future work on dielectric loss, stress engineering, and superconducting circuit integration.

Noise measurements indicate that optically induced microwave heating can be mitigated through pulsed operation and lower optical powers. Extrapolating expected near-term device improvements suggests that high-efficiency conversion at negligible added-noise levels should be achievable on this platform.

Taken together, these results position thin-film lithium tantalate as a scalable, stable, and high-performance electro-optic technology for quantum interconnects. With continued advances in optical loss, microwave coherence, and coupling strength, TFLT transducers could become practical interfaces linking superconducting quantum processors through low-loss optical fiber networks and enabling modular quantum computing at scale.

\paragraph{Acknowledgements}

The authors thank A. Pintart, V. Karam, A. Tusnin, F. Fischer, C. Giroud, D. Pittilini, and A. Feofanov for support in developing Miraex' photonics platform including valuable discussions.
We thank I.C. Benea-Chelmus and HYLAB, V. Manucharyan and SQIL, and Y. Chu and HyQu for sharing resources and expertise.
We thank D. Brau, J. Djukic and A. Joye for their dedication to a supportive environment in which this work was performed.

\printbibliography

\appendix

\renewcommand\thefigure{\thesection.\arabic{figure}}    
\setcounter{figure}{0}
\renewcommand\thetable{\thesection.\arabic{table}}    
\setcounter{table}{0}

\section{Methods: Device}

\subsection{Transducer design}
\label{app:hamiltonian}

The bare electro-optic interaction Hamiltonian describing bi-directional conversion can be expressed as \cite{soltani_efficient_2017}
$$
H_\text{eo} = \hbar g_0 \left( \hat{a}^\dagger_\text{AS} \hat{a}_\text{S} \hat{a}_\text{m} + \hat{a}^\dagger_\text{S} \hat{a}_\text{AS} \hat{a}_\text{m}^\dagger \right)
$$
where $\hat{a}_\text{S,AS,m}$ operators annihilate photons in the S, AS, or m modes respectively, and $g_0$ is the single-photon electro-optic coupling rate.

Strongly driving the S mode with an optical pump with $n_\text{S} = |\alpha_\text{S}|^2$ intracavity photons, $H_\text{eo}$ linearizes to the form of a beamsplitter,
$$
H = \hbar g_0 \sqrt{n_\text{S}} \left( \hat{a}^\dagger_\text{AS} \hat{a}_\text{m} + \hat{a}_\text{AS} \hat{a}^\dagger_\text{m} \right) \text{ .}
$$

The interaction rate is set by the Pockels effect-induced frequency shift caused by the application of microwave field to the lithium tantalate.
It can be derived from first-order perturbation theory, and with interaction dominated by a uniform microwave field (here, $E_{\text{m},z}$ between the capacitors) aligned with the electro-optic tensor element $r_{33}$, $g_0$ takes the form
$$
g_0 = \frac{\epsilon_0 \omega_\text{opt}}{4 U_\text{opt}} \int_\text{EO} r_{33} E_{\text{m},z}^\text{(zpf)} \bf{E}^{*}_\text{AS} \cdot \bf{E}_\text{S} \, dV
$$
for vacuum permittivity $\epsilon_0$, average optical resonator frequency $\omega_\text{opt} \sim \omega_\text{AS} \sim \omega_\text{S}$, optical resonator energy $U_\text{opt}$, $\bf{E}^{*}_\text{S,AS}$ the electric fields of S and AS modes, and $E_{\text{m},z}^\text{(zpf)}$ the microwave zero-point field in the interaction region.
Since the product $\bf{E}^{*}_\text{AS} \cdot \bf{E}_\text{S}$ is odd, $E_{\text{m},z}$ must also be odd to integrate to a non-zero $g_0$.
This leads to the anti-symmetric orientation of capacitor plates shown in Figure~\ref{fig:scheme}(a).

The rate $g_0$ can be further simplified with a geometric factor $\Gamma_\text{eo} = \alpha \xi$ including electrode coverage fraction $\alpha$ and the signed overlap integral of microwave and optical fields $\xi$, as
$$
g_0 \approx \frac{\omega_\text{opt}}{2} n_e^2 r_{33} E_{\text{m},z}^\text{(zpf)} \Gamma_\text{eo} \text{ .}
$$

In our designs, changing the straight length of the resonator changes $E_{\text{m},z}^\text{(zpf)} \Gamma_\text{eo}$ by diluting the field and simultaneously increasing the coverage factor $\alpha$ (within $\Gamma_\text{eo}$) describing the fraction of interaction region relative to the optical resonators' total extent.
These values are shown in Table~\ref{tab:parameters}.

Waveguides are wider in the interaction regions (to maintain low loss) than in the bends and coupling regions (to improve coupling strength while reducing the transfer of energy to higher order modes).
Evanescent coupling between the bright and dark resonators depends the size of gap $s_2$, leading to minimum splitting values 3.4--6.2\,GHz.

Independent measurements of gap-induced optical loss suggest that these dimensions place worst-case limits on propagation loss of \SI{0.12}{dB/\meter} and \SI{0.28}{dB/\meter} in these sections respectively.
In both sections, the electrodes make direct contact with the TFLT slab layer, overlapping by about \SI{1}{\micro\meter}, which should concentrate the microwave electric field to improve $g_0$ by a small factor \cite{warner_coherent_2025}.

As resonator straight length $L$ increases from \SI{150}{\micro\meter} to \SI{500}{\micro\meter} in our designs, the coverage factor increases from $\alpha=0.53$ to $\alpha=0.60$ as the fraction of the resonator formed out of bends decreases.
At the same time, field magnitude $|E_{\text{m},z}|$ reduces about 20\% as capacitance increases and fields become less uniform as microwave resonator size approaches the microwave wavelength, which reduces the level of anti-symmetry and degrades $\Gamma_\text{eo}$.
Including all simulated contributions, our ideal $g_0 / (2 \pi)$ is expected to vary from 1.30--1.03\,kHz between short and long designs.

Cooperativity $C_\text{eo}$ in the limit $C_\text{eo} \ll 1$ can be given by \cite{han_microwave-optical_2021}
$$
C_\text{eo} = \frac{\eta}{4} \frac{\kappa_0^{(\text{b})}}{\kappa_\text{ext}^{(\text{b})}} Q_\text{int} (Q_\text{int} + Q_\text{ext})
$$
in terms of the values given in Table~\ref{tab:parameters}.

Microwave signals are capacitively coupled from a coplanar waveguide (CPW) transmission line as depicted in Figure~\ref{fig:scheme}(c).
Coupling is regulated by the width of the ground plane separating the transmission line and the nearest inner capacitor electrode of the transducer resonator.

\subsection{Cryogenic packaging}
\label{app:packaging}

To package devices for cryogenic measurements, we align fiber array units (FAUs) to grating coupler structures patterned in LT and fix them permanently with UV-cured epoxy.
Transmission level and peak transmission wavelength can change with the application of epoxy, dependent on whether or not the epoxy is filling the optical interface.
Transmission level of packaged devices can also change with temperature.
From calibrated transmission measurements we infer about 6\,dB loss per coupler at room temperature to \SI{9}{dB} or more per coupler when attached and cooled to below \SI{4}{\kelvin}.
The majority of this additional loss is due to diffraction from the FAU before interaction with the grating coupler.
The values of optical resonator loss measured at cryogenic temperatures are generally equal to those measured at room temperature, although a shift in grating coupler wavelength, shift of resonances with temperature, and an increase in coupling loss increasing reflections can all modify loss or modify its wavelength dependence.

\subsection{Optical resonator performance}
\label{app:resonators}

In the data presented in Figure~\ref{fig:bias}(c), we fit to a dual log-normal distribution modeling multiplicative losses from multiple effects, reporting median ($\mu^{*}$) and standard deviation $\sigma$ which is related to a multiplicative scatter parameter as $e^\sigma$.
For ring resonators, we measure resonances between 1556--1559~nm on 52~unique chips from 2~wafers, finding 80\% of resonators' losses ($\kappa_0/2\pi$) falling within a distribution with median $\mu^{*}=121$\,MHz and $\sigma=0.45$, and the remainder in a distribution with $\mu^{*}=59.5$\,MHz and $\sigma=0.14$.
For mid-process transducer resonators, a small sample size of 51~resonances between 1550--1565\,nm from one chip gives narrowly spread data (79\% in $\mu^{*}=177$\,MHz, $\sigma=0.19$; 21\% in $\mu^{*}=148$\,MHz, $\sigma=0.022$).
For post-process transducer resonators, 11~chips from 2~wafers measured between 1550--1564\,nm give 91\% of devices within a primary distribution ($\mu^{*}=229$\,MHz, $\sigma=0.34$) and 9\% of devices within a secondary tail ($\mu^{*}=571$\,MHz, $\sigma=0.16$).

Ring resonators generally have lower loss than transducer resonators at the same mid-process step, despite similar widths in the bends, which could be explained by their compact size making them less likely to encounter scattered debris or spatial process or material variations like film non-uniformity.
Post-process transducer resonators are slightly degraded, perhaps due to metal layer processing or the presence of metal structures or debris, but still tightly clustered.
The secondary distribution is attributed to resonances that have experienced bend-induced mode mixing, affecting resonances at certain wavelengths.
No statistically significant difference is observed between optical resonances measured at room temperature and $<4$\,K.

\subsection{Fabrication}

\label{app:fabrication}

Transducers are fabricated at wafer scale on \SI{400}{\nano\meter} of X-cut lithium tantalate (LT) on top of \SI{2}{\micro\meter} of thermally grown silicon dioxide on top of a high resistivity silicon substrate (NanoLN). 
Optical waveguides are patterned using stepper based DUVL and etched in two steps: first creating the rib waveguides about \SI{200}{\nano\meter} deep, then removing remaining LT from the wafer outside of an area extending about \SI{2}{\micro\meter} from the edge of the rib waveguides.
In this region, about \SI{200}{\nano\meter} of LT slab remains.
Redeposited LT produced by the dry etching process was removed through wet etching. 
A \SI{50}{\nano\meter} NbN film is deposited by magnetron sputtering and patterned using DUVL and a lift-off process.

The optical waveguides are fully ``air cladded'' with no additional materials topping the rib waveguides.
Superconducting bridges bridging electric circuitry across optical waveguides are composed of a Nb layer patterned using liftoff and formed using a photoresist reflow process.
The chips are singulated using blade dicing during the fabrication of the the bridges.

\subsection{Microwave measurements}
\label{app:microwave}

Microwave resonances are measured using a vector network analyzer as a function of microwave power and cryostat temperature, and complex $S_{21}$ transmission is fitted by to a ``hanger'' model that includes an allowance for asymmetric bus coupling \cite{probst_efficient_2015}.
Transducer resonances, particularly those with quality factors below 1,000, are most easily separated from background signal by observing a clear characteristic shift of resonant frequency and $Q_\text{int}$ as a function of temperature, or by performing microwave spectroscopy when attempting optical--microwave transduction.

To assess the performance of the bridges, resonators are produced on devices without transducers using the same bridge process as for transducer devices.
Different numbers of bridges are pattern in series as part of microwave resonators, and the resonators' quality factor is determined.
These measurements show no significant dependence of resonator quality factors on the numbers of bridges in series, suggesting that bridges do not limit internal quality factors, whether from bulk/surface dielectric losses or contact resistance.

\subsection{Microwave participation ratio analysis}
\label{app:participation}

To understand this, we use independent resonator measurements and bulk participation ratio analysis to place bounds on material loss tangents.
We obtain a bound $\tan \delta_\text{LT} < 2.9 \times 10^{-3}$ on the loss tangent of etched TFLT and $\tan \delta_\text{LN} < 5.2 \times 10^{-3}$ on etched TFLN, both measured at cryogenic temperatures.
Our bound on TFLT is more than an order of magnitude worse than what has been reported in literature for bulk LT \cite{jacob_temperature_2004}.
We further place a bound on our Si substrate of $\tan \delta_\text{Si} < 2.1 \times 10^{-6}$ and on the thermal oxide (BOX) of $\tan \delta_{\text{SiO}_2} < 5.1 \times 10^{-5}$.
In our transducers, we simulate 50\% of the energy of the microwave resonant mode participating in the substrate, followed by 30\% in the BOX and 10\% in the TFLT.
These energy participation ratios place individual bounds of $Q_\text{int} > $\,3,500, $Q_\text{int} > $\,65,000, and $Q_\text{int} > $\,600,000 due to our measured TFLT, BOX, and substrate losses respectively.
Using the bulk LT literature bound, we would be limited to $Q_\text{int} > $\,100,000.

\section{Methods: Transduction}

\subsection{Operation of characterization setup}
\label{app:setup}

A microwave tone is optionally amplified and applied to the device for microwave spectroscopy or microwave--optical transduction.
Optical tones are amplified with an erbium doped fiber amplifier (EDFA), gated with an acousto-optic modulator (AOM), and their polarization controlled prior to entering the cryostat.
The transducer is held at a fixed temperature between \SI{0.1}{\kelvin} and \SI{3.8}{\kelvin}, depending on the experiment (see Table~\ref{tab:parameters}).

Microwave and optical frequencies are identified through spectroscopy measurements in each respective domain.
Bias voltage is varied to find the condition where resonator splitting matches the microwave resonance frequency.
Once a transduction signal is obtained, it is used to periodically check and calibrate laser wavelength, polarization, and bias voltage, providing a `slow' feedback loop for the characterization.

Microwave and optical transmission through all circuit stages is independently calibrated so as to be able to extract on-chip transduction efficiency.

Baseline predictions of $g_0$ were obtained by comparing computed fields in our LT designs with those of LN literature designs, accounting for material and design differences, and then scaling $g_0$ values reported in those experiments.
This approach is more likely to underestimate $g_0$ than explicit microwave--optical field simulations, since effects like piezoelectric loss may have lowered $g_0$ in those reports.

While performing repeated transduction efficiency measurements, we monitored experimental system parameters that could affect inferred efficiency such as input light polarization state and the laser frequency.
The bias voltage found to produce maximal transmission efficiency did not generally appear to drift over time, nor did the value of splitting frequency between optical resonances change.
This feedback was primarily applied to correct for laser frequency and polarization changes.

\subsection{Added noise measurement and modeling}
\label{app:noise}

We are using input--output theory to model \cite{holzgrafe_cavity_2020} the incident optical noise as populating a thermal bath, characterized by photon number $N_\text{res}$, that couples to the microwave resonant mode through its internal quality factor $Q_\text{int}$.
At the same time, the microwave mode interacts (through its external coupling factor $Q_\text{ext}$) with the microwave bus with characteristic bath population $N_\text{wg}$.
Both $N_\text{res}$ and $N_\text{wg}$ contribute to the total output added noise $N_\text{add,out}$ that leaves the chip in the direction of the detector.
Notably, since the noise is not added by the physics of the transduction process itself, but rather the light--superconductor interaction, values of $N_\text{add,out}$ measured this way should be equivalent to input-referred added noise as typically referenced in \emph{microwave--optical} transduction experiments (where the added noise undergoes conversion into the optical domain for detection, with sub-unity gain). 

\begin{figure}[htbp]
    \centering
    \includegraphics[width=1\linewidth]{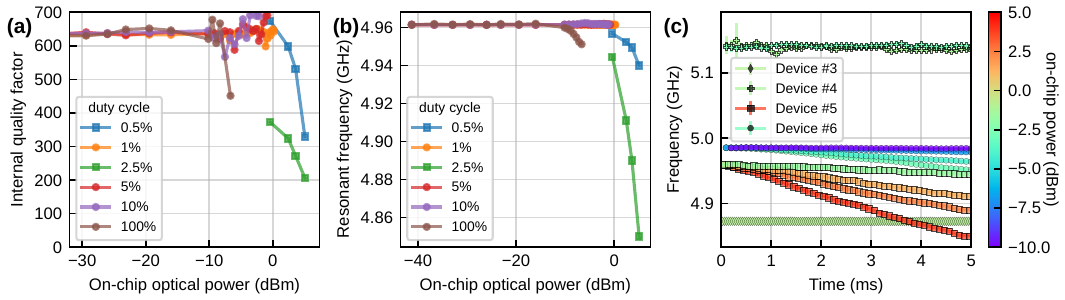}
    \caption{
    Microwave resonator response to applied optical power.
    (a-b) On Device~5, optical pulses with different pulse length, repetition rate, and peak powers are applied.
    The time-averaged microwave transmission is measured and microwave resonance parameters are fit to obtain (a) internal quality factor $Q_\text{int}$ and (b) resonance frequency, which both shift as total applied optical energy increases.
    Pulses are longer than the dynamics we typically observe at low powers and duty cycle, justifying time averaging.
    For data taken at high optical powers (squares), pulse period is fixed at \SI{200}{\milli\second} and pulse width is varied from \SI{1}{\milli\second} to \SI{5}{\milli\second}.
    For data at lower powers (circles), pulse width is fixed at \SI{50}{\milli\second} and the pulse period is varied from \SI{0.5}{\second} to \SI{5}{\second}.
    (c) Time dynamics of resonance frequency within a pulse for different optical powers on several devices, obtained by time-resolved spectroscopy measurements and accompanying the data in Figure~\ref{fig:noise}(b).
}
    \label{fig:shifts}
\end{figure}

We perform optical-to-microwave transduction and noise measurements with duty cycles 0.005--5\%, pulse widths 0.01--10\,ms, and peak cryogenic (on-chip) optical powers up to about \SI{10}{\milli\watt} (\SI{1}{\milli\watt}) (Figure~\ref{fig:shifts}).
For this range of configurations, we generally do not observe a change in the background noise level prior to each pulse as compared to the case with no power (0\% duty cycle), indicating that the chip is fully thermalized between pulses.

\begin{table*}[ht]
\centering
\caption{Summary of device parameters and measured properties.
Geometrical parameters associated with varied design properties are shown in Figure~\ref{fig:scheme}.
``Tunability'' refers to the shift of dark resonator frequency with respect to the applied bias voltage.
}
\label{tab:parameters}
\begin{tabular}{lcccccc}
\toprule
\textbf{Device \#} & 1 & 2 & 3 & 4 & 5 & 6 \\
\midrule
\textbf{Wafer \#} & 1 & 1 & 1 & 2 & 2 & 2 \\
\midrule
\multicolumn{1}{c}{\textbf{Design properties}} \\
\midrule
Straight length $L$ (µm) & 500 & 400 & 350 & 400 & 350 & 350 \\
Bus gap $s_1$ (nm) & 646 & 663 & 672 & 663 & 672 & 672 \\
Coupling gap $s_2$ (nm) & 410 & 410 & 410 & 470 & 470 & 470 \\
Minimum splitting (GHz) & 4.03 & 4.68 & 4.73 & 4.06 & 4.40 & 4.23 \\
Free spectral range (GHz) & 62 & 69 & 73 & 69 & 73 & 73 \\
Coverage fraction $\alpha$ & 0.60 & 0.58 & 0.58 & 0.58 & 0.58 & 0.58 \\
Coverage--field density\\\,\,\,\,\,product $\alpha |E_{\text{m},z}|$ & 1.60 & 1.72 & 1.84 & 1.72 & 1.84 & 1.84 \\
\midrule
\multicolumn{1}{c}{\textbf{Electro-optic response}} \\
\midrule
Tunability\\\,\,\,\,\,warm (GHz/V) 
& $0.650(3)$ & $0.592(6)$ & $0.53(3)$ 
& $0.62(3)$ & $0.67(2)$ & $0.72(1)$ \\
Tunability\\\,\,\,\,\,cold (GHz/V) 
& $0.315(4)$ & $0.179(4)$ & $0.301(7)$ 
& $0.25(1)$ & $0.364(2)$ & $0.337(5)$ \\
Inferred $g_0/(2\pi)$ (Hz)
& 1130 & 480 & 280 
& 1150 & 54 & 150 \\
Cooperativity $C_\text{eo}$
& 0.011 & 0.0022 & 0.0078
& 0.021 & $6.2 \times 10^{-5}$ & 0.0015 \\
\midrule
\multicolumn{1}{c}{\textbf{Microwave properties}} \\
\midrule
Resonant\\\,\,\,\,\,frequency $\omega_\text{m}$ (GHz) 
& 5.538 & 5.285 & 4.873 & 5.140 & 4.961 & 4.985 \\
$Q_\text{int}$ 
& $265(10)$ & $420(100)$ & $1000(100)$ 
& $600(100)$ & $620(100)$ & $900(100)$ \\
$Q_\text{ext}$ 
& $850(10)$ & $180(100)$ & $7500(500)$ 
& $12000(500)$ & $6000(1000)$ & $3500(500)$ \\
\midrule
\multicolumn{1}{c}{\textbf{Optical properties}} \\
\midrule
$\kappa_0^{(\text{r})}$ (MHz) 
& $203(1)$ & $217(7)$ & $164(1)$ 
& $359(3)$ & $290(2)$ & $235(1)$ \\
$\kappa_0^{(\text{b})}$ (MHz) 
& $138(1)$ & $254(4)$ & $194(1)$ 
& $201(3)$ & $286(1)$ & $342(2)$ \\
$\kappa_\text{ext}^{(\text{r})}$ (MHz) 
& $64.1(3)$ & $19(1)$ & $62(1)$ 
& $94(1)$ & $169(1)$ & $90(1)$ \\
$\kappa_\text{ext}^{(\text{b})}$ (MHz) 
& $15.5(1)$ & $49(1)$ & $38(1)$ 
& $19(1)$ & $41(1)$ & $31(1)$ \\
Transmission loss\\\,\,\,\,\,per coupler (dB) 
& 9.2 & 10.4 & $<9$ & 10.2 & 10.0 & 11.6 \\
\midrule
\multicolumn{1}{c}{\textbf{Operating conditions}} \\
\midrule
Optical wavelength (nm) 
& 1560.70 & 1562.74 & 1562.00 
& 1560.77 & 1560.92 & 1560.96 \\
Device temperature (K) 
& 2.8 & $<1$ & $<1$ & $<1$ & $<1$ & $<1$ \\
\bottomrule
\end{tabular}
\end{table*}

\end{document}